\documentstyle[11pt,newpasp,twoside]{article}
\def\edcomment#1{\iffalse\marginpar{\raggedright\sl#1\/}\else\relax\fi}
\reversemarginpar

\begin{document}
\title{Stellar abundance patterns. \\
 What is the possible level of completeness today ?}
 \author{Gopka V.}
\affil{ Odessa Astronomical observatory (OAO), Odessa National University,  park Shevchenko,
   Odessa, 65014, Ukraine}
\author{Yushchenko A.}
\affil{Chonbuk National University, Chonju, 561-756, (South) Korea \\
       OAO }
\author{Musaev F., Galazutdinov G.}
\affil{The International Centre for Astronomical, Medical and Ecological
       Research of the Russian Academy of Sciences (RAS) and
       the National Academy of Sciences of Ukraine (NAS) -- ICAMER,
       Golosiiv, Kiev, 03680, Ukraine  \\
        Special Astrophysical observatory (RAS),
        Nizhnij Arkhyz, Zelenchuk, Karachaevo-Cherkesiya, 369167, Russia -- SAO RAS}
\author{Kim C.}
\affil {Chonbuk National University, Chonju, 561-756, (South) Korea, }
\author{Shavrina A., Pavlenko Y.}
\affil{Main  Astronomical observatory (NAS),
       Golosiiv, Kiev, 03680, Ukraine }
\author{Polosukhina N.}
\affil{Crimean Astrophysical observatory,
       Nauchnyi, Bakhchisaraj, Crimea, 333413, Ukraine }
\author{North P.}
\affil{Inst. d'Astronomie, Dorigny, CH-1015, Univ. Lausanne,  Switzerland }

\begin{abstract}
%We made a brief review of several papers with the most complete stellar
%abundance samples.
 We discuss the way of increasing of the number of chemical elements,
 investigated in stellar spectra.
 We can reach it by using spectrum synthesis method,
 new atomic data and observation of stellar spectra with resolution
 comparable to solar spectral atlases.
 We show two examples of this kind researches. The first is the
 implementation of new atomic data to well known Przybylski's star.
 We show that the number of spectral lines,
 which can be identificated in the spectrum of this star can be significantly
 higher.
 The second example is the investigation of $\zeta$ Cyg.
 We found the abundances of 51 elements in the atmosphere of
 this mild barium star.
\end{abstract}


\section{Introduction}

     If we will try to overview the results of determinations of stellar
     abundances, we will find that a lot of elements with Z$>$30 are omitted.
     Elements heavier than iron group are synthesized primarily by means
     of neutron capture processes. According to the work by Burbidge et al.
     (1957,  B2FH),  in  order	to  reconstruct  the  Galactic evolutionary
     history  of heavy elements , one  has to consider two major mechanisms
     of   neutron  addition:  the   s-process and
     r-process. Small  amount  of  nuclei can be
     created  in  p-process.  New process  of  heavy  elements creation was
     proposed  by  Woosley \& Hoffman (1992)  -  this is $\alpha$-process for
     heavy   elements.       A  review of the developments
     of  the  theory  of  elements creation  was made by
     Wallerstein  et  al.  (1997).  This  paper  was  devoted  to  the 40th
     anniversary  of  B2FH work.  The  theory of element creation  needs  a
     detailed  abundance pattern  for comparison of theoretical predictions
     with observations.  New  data  on   stellar   abundance  patterns  for
     stars of different types can  significantly influence the theory.

	The  full isotopic pattern is available only for meteoritic matter.
     The  precision  of abundance determinations in  stars  is not so high.
     The best abundance samples for the stars are:

      ~~~~Sun         -- 73 elements (Grevesse \& Sauval, 1998)

      ~~~~Procyon~A   -- 55 elements (Yushchenko \& Gopka, 1996a,b),

      ~~~~Przybylski's star  -- 54 elements  (Cowley et al., 2000),

      ~~~~$\chi$ Lupi -- 51 elements  (Leckrone et al., 1999),

      Gopka (2000) investigated the abundances of several heavy elements
     if the atmosphere of Sirius~A.
       Now we are preparing a review of chemical composition of Sirius~A.
     The abundance pattern of this star consists of 50 elements
       We can see a significant progress. In 1988 Reinolds et al. pointed,
     that the abundance  pattern for Canopus (38 elements) is the third
     after the Sun and  Przybylski's star  (Wegner \& Petford, 1974 --
     51 elements).

     We realize, that this review is not full, but it is sufficient to 
     make some conclusions. 
     First of all -  detailed abundance determination is a very  long process.
     In many cases it cost a decades of efforts of many scientists.
     The majority of above cited papers are the review papers or the papers
     which are the
     final paper in a long series of articles on abundances in this star.         
     What are the necessary conditions for future progress ?
     The  base  of  the improvement  are:

        1) the  increasing  of  level/noise ratio and spectral resolution
           stellar spectroscopy, investigation  of  the  chemical
           abundance  using  UV spectra;

        2) the new  atomic  data;

        3) the using  of the spectrum synthesis method not only for limited
           number of lines in the spectrum, but for majority of lines,
           taking in account hyperfine and isotopic splitting,
           magnetic fields, spots, detailed analysis of spectral binaries,
           individual atmosphere models,  etc.

     Let us show a short overview for these three items.
     We will show only
     some aspects which, we hope, will be interesting for the community
     of this conference. It is difficult to point poor better observations
     or poor better atomic data or software influence on final results.
     Usually it is the combination of all three items.

\section{High quality observations}

     The majority of cited papers, which give us the most complete
     stellar abundance samples, deals with the spectra obtained at the
     best telescopes and spectrographs. But it should be noted, that
     we have 2.5 meter telescope since 1917, 5 meter - since 1948, 6 meter -
     since 1975, 10 meters - since 1991. But while this telescopes were
     the single in it's class, no significant progress were available.
     Only during last two decades, when the number of telescopes
     over 2 meters became near 50, we can see a lot of new
     results.

       During the last decade observations with coude-echelle spectrographs
     were started at the 1 meter telescope of SAO RAS and
     2 meter telescope of ICAMER. The last telescope is located at 3100 meters
     elevation. Now it is possible to
     obtain spectra with resolutions 45000, 80000, 125000, 190000.
     Now we are testing
     the new spectrograph which permit us to obtain the spectra of
     bright stars with the resolution comparable with solar spectral atlases.

       In the last section of this paper we review the
     preliminary  results of one of the
     first detailed investigations of stellar abundances with this
     spectrograph. The  mild barium star $\zeta$ Cyg was observed in 2000
     with relatively low resolution - near 80000.
     This investigation
     will be the base for future observation of this and other barium
     stars with highest spectral resolution in visual and near UV wavelengths
     regions.

\section{New atomic data -- line identification in Przybylski's star}

     The more detailed review of used observations and methodics one can
     find near this poster in the poster of Shavrina et al. (2002).
     Here we will point only important items:

      1) wavelength coverage 6123-6175 \AA\AA ~and 6676-6732 \AA\AA~ -- 108 \AA;

      2) spectral resolution R=100000, signal to noise ratio S/N$>$100;

      3) new lanthanides lines from DREAM database (Biemont et al., 2002);

      4) identification of lines and calculation of abundances were made
     using spectrum synthesis method. We used Kurucz (1995) SYNTHE program
     for calculation synthetic spectra and Yushchenko (1998) URAN program
     for line identification and spectrum synthesis in automatic mode.
     Abundances for all identified lines were found with spectrum synthesis
     method.

     In tables 1,2 we give the summary of results. In table 1 one can find
     the total number of lines of  different elements and ions in
     DREAM database. The numbers in the last column of this table
     -  the part of these lines, which were present in our
    previous line list.
    The total number of lanthanides and thorium
    lines in DREAM database is 56150.

    We tried to use these lines to find the abundances in Przybylski's
    star. We used 108 \AA~~ wavelength interval in red part of the spectrum.
    In table 2 we show our mean abundances, errors and
    number of lines from 108 \AA~ interval in red region of the spectrum
    and the corresponding values from  2693 \AA~  interval
    in blue-red parts of the spectrum according to Cowley et al. (2000).

    The brief inspection of the abundances show, that the results
    are similar in the range of uncertainties.
    In the last column of this table we give the ratio of lines,
    identified in the work of Cowley et al (2000) and in this work.

    From the ratio of wavelengths coverage in two studies we can expect that
    the ratio of numbers of identified lines will be near 25 or more,
    if we take in account, that the majority of lanthanides lines are located
    in the blue part of the spectrum.
    But only one value in the last column of the table exceeds 10.

    It means that the lanthanides lines from DREAM database give us a
    powerful instrument for examination of the atmospheres of peculiar stars.

    It should be noted that not only lanthanides show the overabundance
    in the atmosphere of  Przybylski's star, but all heavy elements.
    Unidentified lines can be the lines of other heavy elements.

\section{Chemical composition of $\zeta$ Cyg}

     $\zeta$ Cyg is one of the brightest middle
     barium stars. The spectrum was obtained at ICAMER  2 meter
     telescope  with resolving power R=80000 in the wavelength range
     3495-10000~\AA~~ with signal to noise ratio in visual and infrared
     region more than 100.
     As a first step we tried to find atmosphere parameters.
     We tested the parameters used by Zacs (1994) and Boyarchyk et al. (2001).
     We analyzed iron lines in the spectrum and found that following
     atmosphere parameters are valid:

     T$_{eff}$=5050~K, $lgg$=2.8, $v_{micro}$=1.45 km/s, $v_{macro}$=3-3.5 km/s,

     We interpolated Kurucz (1995) atmosphere model with these parameters
     and calculated synthetic spectrum for all observed region.
     This spectrum was used for identification of spectral lines.
     Abundance calculations for all elements except iron were made
     with spectrum synthesis method. We used Kurucz (1995) SYNTHE program
     for calculation of synthetic spectra and Yushchenko (1998) URAN program
     for approximation of the observed program by calculated one
     in automatic mode.

     We tried to obtain abundances by direct comparison with the solar
     spectrum. We found solar oscillator strengths for majority of
     the used lines. We used Liege Solar atlas (Delbouille et al., 1974),
     Holweger-Muller atmosphere model,
     microturbulent velocity 1.0 km/s, macroturbulent velocity 1.8 km/s.
     SYNTHE and URAN codes were used for
     approximation of observed solar spectrum by synthetic one.

     In tables 3,4 we show the results of abundance determinations.
     The results of Gratton (1985), Zacs (1994), Boyarchuk et al.(2001)
     are displayed for comparison. The chemical elements with lines,
     that have no counterparts in the solar spectrum are marked by asterisk.

     We found the abundances of 48 elements in the atmosphere of $\zeta$~Cyg.
     The abundances of Li, N, Ba are known from previous investigations
     of this star.
     The total abundance sample consist of 51 elements.
     Detailed analysis of these results will be made later.

\section{Conclusion}

     We show that the abundances of 50-55 elements are the best results
    for stellar abundance samples in sharp-lined stars.
    We can point some ways how to  increase these number
    for several types of the stars.
    First of all F stars of main sequence. The detailed
    chemical composition  of Procyon is a summary of many
    papers, and significant part of this papers were used Griffin's
    (1979) atlas of this star. New atlas of the star observed with
    higher signal to noise ratio and spectral resolution will help to find
    several new elements.

    The other way is the observations of UV spectrum of Procyon or
    Procyon type stars. We have no  determinations of chemical
    composition of Procyon from UV spectral data.

    For barium stars additional abundances can be obtained from
    observations in the near UV wavelengths region - 3100-3500~\AA.

    And let we will not forget about the Sun.
    There are a dozen of chemical elements  with unknown
    abundances in the solar atmosphere. For example Gopka et al. (2001)
    found the abundance of arsenic in the Sun.
    The spectral atlases of solar spots can be used for abundance
    determinations, if the good model of the spot will be made.

    It should be noted that all these ways will lead us to very
    crowded spectral regions. Usual model atmosphere  method is not
    available in this case. It is necessary  to use spectrum synthesis
    and automatic spectrum synthesis.
    The programs for automatic spectrum synthesis were
    described by Cowley (1995), Valenty and Piskunov (1995),
    Tsymbal \& Cowley (2000), Erspamer \& North (2002),
    Yushchenko (1998). These programs are very different,
    but all of them can help us to obtain detailed abundance samples
    for different type stars.

    Here before we have showed the significance of new atomic data for
    lanthanides. But it should be noted that  Corliss \& Bozman (1961)
    oscillator strengths are still in use for several elements.
    We have zero or very limited  information about the abundances
    of elements with Z=50-55, 33-36. One of the peaks of solar system
    abundances is near Z=50-55. And we have zero information about
    this peak in the Sun and in other stars. One of the first attempts to find
    the abundance of tellurium (Z=52) (Yushchenko \& Gopka, 1996a,b)
    lead us to possible overabundance of this element in Procyon.

    Maybe a decade later, when 50-70 elements will be a common number
    of elements
    in every paper on chemical abundance in sharp-lined stars,
    we will be able  to resolve a lot of modern problems in this field
    and to ask new questions.

\acknowledgments

This work was supported by the grant of Post-Doc Program, Chonbuk National 
University, (South) Korea (2002).

\pagebreak

\begin{table}
\caption{Number of lines of different elements in DREAM database}
\centerline{
\begin{tabular}{lrr}
\hline
                 & \multicolumn{2}{c}{Number of lines} \\
                   \cline{2-3}
        Element  &    in Dream   &   which are common   \\
                 &               & in Dream and in      \\
                 &               & our old line list        \\
\hline
             La III    &     137 &           9   \\
             Ce II     &   14970 &        1663   \\
             Pr III    &   18401 &           -   \\
             Nd III    &      51 &          51     \\
             Tb III    &     923 &           -       \\
             Ho III   &     1324 &           -      \\
             Er III   &     1307 &         304        \\
             Tm II    &     7954  &        474          \\
             Tm III   &     1478  &          -            \\
             Yb II    &     5484  &        311              \\
             Yb III   &      278  &          -   \\
             Yb IV    &     2769  &          -   \\
             Lu II    &      106  &         74     \\
             Lu III   &       58  &          3       \\
             Th III   &      901  &          -         \\
\hline
\end{tabular}
  }
\end{table}

\begin{table}
\caption{Abundances of lanthanides in Przybylski's star}
\centerline{
\begin{tabular}{ l rrr rrr r }
\hline
                    & \multicolumn{3}{c}{Results with}&\multicolumn{3}{c}{Cowley et al., 2000,}&  \\
                    & \multicolumn{3}{c}{DREAM lines }&\multicolumn{3}{c}{  MNRAS, 317, 299}   &  \\
\cline{2-8}
          Wavelengths& \multicolumn{3}{c}{6123-6175 = 52~\AA}&\multicolumn{3}{c}{3959-6652 = 2693~\AA} & \\
          interval  & \multicolumn{3}{c}{6676-6732 = 56~\AA}&                  &   &   &   \\
\cline{5-8}
                    &  &   &    &       \multicolumn{4}{r}{2693/(52+56) = 25} \\
\hline
            Element &   lgN & $\sigma$& N1 &     lgN  & $\sigma$ &  N2 &  N2/N1 \\
\hline
            La II   &  -8.32& 0.25 &  4 &    -8.17 & 0.29&  28 &   7    \\
            Ce II   &  -7.58& 0.06 & 28 &    -7.60 & 0.26&  46 &  1.6   \\
            Pr I    &       &      &    &    -6.40 & 0.21&   4 &        \\
            Pr II   &  -9.45& 0.13 &  3 &    -8.80 & 0.21&  31 &  10    \\
            Pr III  &  -8.25& 0.52 &  3 &    -7.46 & 0.16&  12 &   4    \\
            Nd I    &       &      &    &    -6.39 & 0.35&   6 &        \\
            Nd II   &  -7.62& 0.09 & 10 &    -7.65 & 0.28&  71 &   7    \\
            Nd III  &  -6.89&      &  1 &    -7.31 & 0.30&   7 &   7    \\
            Sm II   &  -7.65& 0.07 & 10 &    -7.75 & 0.29&  41 &   4    \\
            Eu II   &  -9.15&      &  1 &    -8.58 & 0.19&   5 &   5    \\
            Gd II   &  -7.61&      &  2 &    -7.62 & 0.25&  35 &  18    \\
            Tb II   &  -8.84&      &  1 &    -8.89 & 0.16&   3 &   3    \\
            Dy II   &  -7.68&      &  2 &    -7.88 & 0.23&  16 &   8    \\
            Ho I    &       &      &    &    -6.60 &     &   1 &        \\
            Er I    &  -6.24&      &  1 &          &     &     &        \\
            Er II   &  -8.17&      &  2 &    -8.09 & 0.22&  18 &   9    \\
            Er III  &       &      &    &    -6.83 & 0.04&   4 &        \\
            Tm II   &       &      &    &    -8.20 & 0.28&  15 &        \\
            Yb II   &       &      &    &    -8.99 & 0.33&   9 &        \\
            Lu II   &  -8.80&      &  1 &    -8.65 & 0.14&   6 &   6    \\
\hline
\end{tabular}
  }
\end{table}

\begin{table}
\caption{Chemical composition of $\zeta$ Cyg}
\begin{tabular}{rr r rrr c rrr l}
\hline
    &	 &Boyar-&\multicolumn{3}{c}{Zacs, 1994}
				     &Gratton&\multicolumn{3}{c}{This work}&  \\
  n &  Z &chuk~~   & $*$ - $\odot$&$\sigma$ & N   &  1985 & $*$ - $\odot$&$\sigma$ & N& Element  \\
    &	 &et al.~  &	  &	 &   &	     &	       &      &  &	    \\
    &	 & 2001~~  &	  &	 &   &	     &	       &      &  &	    \\
\hline
  1 &  3 &	   &-0.14 &	 &  1&	     &	       &      &  &   Li I   \\
  2 &  6 &         &  -   &      &   &  -0.10&   -0.00 & 0.21 & 4&   C  I   \\
  3 &  7 &       ÿ &  -   &      &   &  +0.61&         &      &  &   N  I   \\
  4 &  8 &         &  -   &      &   &  -0.34&   -0.46 &      & 1&   O  I   \\
  5 & 11 &  +0.19  &-0.35 &	 &  2&	     &	 -0.37 & 0.16 & 5&   Na I   \\
  6 & 12 &	   &-0.51 &	 &  2&	     &	 +0.22 & 0.22 & 6&   Mg I   \\
    &	 &	   &	  &	 &   &	     &	 +0.26 &      & 2&   Mg II  \\
  7 & 13 &  +0.16  &	  &	 &   &	     &	 -0.11 & 0.06 & 8&   Al I   \\
  8 & 14 &  +0.09  &+0.14 &  0.13&  3&	     &	 -0.05 & 0.13 &52&   Si I   \\
  9 & 15 &	   &	  &	 &   &	     &	 +0.10 &      & 1&   P	I   \\
 10 & 16 &	   &	  &	 &   &	     &	 +0.20 & 0.26 & 5&   S	I   \\
 11 & 19 &	   &  -   &	 &   &	     &	 +0.00 & 0.26 & 3&   K	I   \\
 12 & 20 &  -0.03  &+0.08 &  0.19&  4&	     &	 +0.02 & 0.09 & 5&   Ca I   \\
 13 & 21 &  -0.02  &+0.04 &  0.30&  4&	     &	 +0.06 & 0.11 & 4&   Sc I   \\
    &	 &	   &	  &	 &   &	     &	 +0.08 & 0.15 &10&   Sc II  \\
 14 & 22 &  -0.11  &-0.17 &  0.22& 21&	     &	 -0.07 & 0.06 &23&   Ti I   \\
    &	 &	   &	  &	 &   &	     &	 +0.00 & 0.11 &31&   Ti II  \\
 15 & 23 &  -0.04  &-0.13 &  0.18& 16&	     &	 -0.01 & 0.12 &40&   V	I   \\
 16 & 24 &  -0.08  &-0.06 &  0.18& 11&	     &	 -0.11 & 0.06 &26&   Cr I   \\
    &	 &	   &	  &	 &   &	     &	 +0.14 & 0.06 &14&   Cr II  \\
 17 & 25 &	   &-0.30 &  0.13&  5&	     &	 -0.26 & 0.15 &21&   Mn I   \\
 18 & 26 &  -0.03  &+0.12 &  0.23& 51&	     &	 +0.02 & 0.10 &92&   Fe I   \\
    &	 &	   &	  &	 &   &	     &	 +0.06 & 0.08 & 6&   Fe II  \\
 19 & 27 &  -0.13  &-0.22 &  0.12&  6&	     &	 -0.02 & 0.09 &11&   Co I   \\
 20 & 28 &  -0.09  &-0.05 &  0.25& 10&	     &	 +0.04 & 0.07 &28&   Ni I   \\
 21 & 29 &	   &	  &	 &   &	     &	 +0.44 & 0.45 & 3&   Cu I   \\
 22 & 30 &	   &	  &	 &   &	     &	 +0.00 & 0.24 & 4&   Zn I   \\
 23 & 32 &	   &	  &	 &   &	     &	 +0.28 &      & 1&   Ge I   \\
 24 & 37 &	   &	  &	 &   &	     &	 -0.12 &      & 1&   Rb I   \\
 25 & 38 &	   &	  &	 &   &	     &	 +0.22 & 0.10 & 3&   Sr I   \\
 26 & 39 &  +0.30  &+0.37 &  0.24&  3&	     &	 +0.15 & 0.15 & 3&   Y	I   \\
    &	 &	   &	  &	 &   &	     &	 +0.48 & 0.16 &22&   Y	II  \\
 27 & 40 &	   &-0.08 &  0.20&  5&	     &	 +0.24 & 0.04 & 8&   Zr I   \\
    &	 &	   &	  &	 &   &	     &	 +0.61 & 0.11 & 7&   Zr II  \\
 28 & 41 &	   &	  &	 &   &	     &	 +0.13 &      & 1&   Nb I  *\\
 29 & 42 &	   &	  &	 &   &	     &	 +0.13 & 0.12 & 3&   Mo I   \\
 30 & 44 &	   &	  &	 &   &	     &	 -0.02 &      & 2&   Ru I   \\
 31 & 45 &	   &	  &	 &   &	     &$<$+0.2~ &      & 2&   Rh I  *\\
 32 & 46 &	   &	  &	 &   &	     &	 +0.36 &      & 1&   Pd I  *\\
 33 & 49 &	   &	  &	 &   &	     &	 -0.12 &      & 1&   In I   \\
\hline
\end{tabular}
\end{table}

\begin{table}
\caption{Chemical composition of $\zeta$ Cyg. Continuation }
\begin{tabular}{rr r rrr c rrr l}
\hline
    &	 &Boyar-&\multicolumn{3}{c}{Zacs, 1994}
				     &Gratton&\multicolumn{3}{c}{This work}&  \\
  n &  Z &chuk~~   & $*$ - $\odot$&$\sigma$ & N   &  1985 & $*$ - $\odot$&$\sigma$ & N& Element  \\
    &	 &et al.~  &	  &	 &   &	     &	       &      &  &	    \\
    &	 & 2001~~  &	  &	 &   &	     &	       &      &  &	    \\
\hline
 34 & 56 &  +0.54  &+0.41 &  0.13&  3&	     &	       &      &  &   Ba II  \\
 35 & 57 &  +0.45  &+0.38 &  0.15&  3&	     &	 +0.51 & 0.20 &12&   La II  \\
 36 & 58 &  +0.33  &+0.55 &	 &  1&	     &	 +0.36 & 0.16 &43&   Ce II  \\
 37 & 59 &  +0.43  &+0.32 &  0.13&  3&	     &	 +0.19 & 0.19 & 6&   Pr II  \\
 38 & 60 &  +0.23  &	  &	 &   &	     &	 +0.42 & 0.17 &70&   Nd II  \\
 39 & 62 &	   &	  &	 &   &	     &	 +0.31 & 0.15 &14&   Sm II  \\
 40 & 63 &  +0.22  &-0.05 &	 &  2&	     &	 +0.45 & 0.05 & 4&   Eu II  \\
 41 & 64 &	   &	  &	 &   &	     &	 +0.27 & 0.19 & 4&   Gd II  \\
 42 & 65 &	   &	  &	 &   &	     &	 +0.12 &      & 1&   Tb II *\\
 43 & 66 &	   &	  &	 &   &	     &	 +0.28 & 0.19 & 5&   Dy II  \\
 44 & 68 &	   &	  &	 &   &	     &	 +0.35 &      & 1&   Er II  \\
 45 & 69 &	   &	  &	 &   &	     &$<$+0.2~ &      & 1&   Tm II  \\
 46 & 72 &	   &	  &	 &   &	     &	 +0.45 &      & 1&   Hf II *\\
 47 & 76 &	   &	  &	 &   &	     &	 +0.30 &      & 2&   Os I  *\\
 48 & 77 &	   &	  &	 &   &	     &$<$+0.5~ &      & 2&   Ir I  *\\
 49 & 78 &	   &	  &	 &   &	     &$<$+0.5~ &      & 1&   Pt 1  *\\
 50 & 81 &	   &	  &	 &   &	     &$<$+0.5~ &      & 1&   Tl I  *\\
 51 & 82 &	   &	  &	 &   &	     &$<$+0.2~ &      & 2&   Pb I  *\\
\hline
\end{tabular}
\end{table}

\end{document}